\documentclass[twocolumn,showpacs,preprintnumbers,amsmath,amssymb]{revtex4}

\usepackage{graphicx}
\usepackage{dcolumn}
\usepackage{bm}

\usepackage{epsfig} \usepackage{changebar} \usepackage{amsmath}

\newcommand {\kv}{\langle k \rangle}

\newcommand{\be}{\begin{equation}}
\newcommand{\ee}{\end{equation}}
\newcommand{\bea}{\begin{eqnarray}}
\newcommand{\eea}{\end{eqnarray}}

\begin{document}

\bibliographystyle{plain}

\preprint{}

\title{Worm Epidemics in Wireless Adhoc Networks}

\author{Maziar Nekovee}

\affiliation{
BT Research, Polaris 134, Adastral Park, Martlesham, Suffolk IP5 3RE, UK \\
and \\
Centre for Computational Science, University College London, 20 Gordon Street, London WC1H 0AJ, UK}


\begin{abstract}
A dramatic increase in the number of computing 
devices with wireless communication capability has 
resulted in the emergence of a new class of  
computer worms which specifically 
target such devices. The most striking feature of 
these worms is that they do not require Internet connectivity for their 
propagation but can spread directly from device to device using
a short-range radio communication technology, such as WiFi or Bluetooth.
In this paper we develop a new model for epidemic spreading 
of these worms and investigate their spreading in wireless adhoc networks via 
extensive Monte Carlo simulations. Our studies show that the threshold 
behaviour and dynamics of worm epidemics in these networks are greatly 
affected by a combination of spatial and temporal correlations which 
characterise these networks, and are significantly
different from the previously studied epidemics in the Internet.
 
\end{abstract}

\pacs{89.75.Hc,05.70.Jk,87.19.Xx, 89.75.Fb}

\maketitle

\section{INTRODUCTION}
Worms are self-replicating computer viruses which can propagate through 
computer networks without any human intervention 
\cite{worm1, worm2, book1-internet}. Cyber attacks by this type of viruses  
present one of the most  dangerous threats to the
security and integrity of computer and telecommunications
networks. The Code Red \cite{codered, book-virus} and Nimda \cite{book-virus}
worms, for example, infected  hundreds of thousands of computers 
at  alarming speeds and the resulting worm epidemics 
cost both the public and the private sector a great deal of money.  
The last few years have  seen the emergence of a new type of worms 
which specifically targets portable computing devices, such as smartphones 
and laptops. The novel feature of these 
worms is that they do not necessarily require Internet connectivity 
for their propagation. They  can 
spread  directly from device to device using a short-range 
wireless communication technology, such as WiFi 
or Bluetooth \cite{book-virus, blue-worm, wi-worm, sciam}, 
creating in their wake an 
{\em adhoc} contact network along which they propagate.
The first computer worm written specially for 
wireless devices was detected in 2003 and within three years the number of 
such viruses soared from one to more than $300$ \cite{sciam}.
With wireless networks  becoming increasingly popular, many 
security experts predict that these networks will soon be a main 
 target of attacks by worms and other type of malware.
\cite{sciam}.

Worm and virus attacks on the Internet have been the subject of extensive 
empirical, theoretical and simulation studies
\cite{worm1,sis1_vesp, email_newman, immune1, immune2}. 
These studies have 
greatly contributed to our understanding of the impact of network 
topology on the properties of virus spreading \cite{sis1_vesp, email_newman}
and have inspired the design of more effective 
immunisation strategies to prevent and combat Internet
epidemics \cite{immune1, immune2}. 
Investigation of virus spreading in 
wireless networks in general and worms in particular is, however,
at its infancy, and there has been very limited studies which address this 
problem \cite{wi-worm, maz1}.

In this paper we develop a new model for the spreading of  
worms in Wi-Fi-based wireless adhoc networks and investigate 
the properties of worm epidemics in these networks via extensive 
Monte Carlo simulations. Wireless adhoc networks \cite{ramin, geiner1, geiner2, geiner2, geiner3, geiner4}
are distributed networks which can be formed on the fly
by WiFi-equipped devices, such as laptops and smartphones.
Nodes in these networks communicate directly with each other 
and can route  data packets wirelessly, 
either among themselves or to the nearest Internet accesspoint.
Adhoc technology has important applications in the 
provisioning of ubiquitous wireless Internet access, 
disaster relief operations and wireless sensor networks. 
From the perspective of complex network theory \cite{review_albert, 
review_newman, review_mendes, review_yamir} 
the study of these networks is important as their topology 
provides a clear-cut example of spatial networks \cite{spat1}.
Spatial networks are embedded in a metric space where interactions 
between the nodes is a function of their spatial distance \cite{spat1, spat2}.
Despite their relevance to many real-life phenomena the properties of these 
networks are much less studied than abstract graphs.

Our Monte Carlo simulations show that epidemic spreading in wireless 
adhoc networks is significantly  
different from the previously studied epidemics in 
the Internet. The initial growth of the epidemic is 
significantly slower than the exponential growth observed for  
worm spreading in the Internet, and  the epidemic prevalence exhibits 
a density-dependent critical threshold which is higher than the value 
predicted by the mean-field theory. We show that these differences are due 
to strong spatial and temporal correlations which characterise these networks.
Our study also reveals the presence of a self-throttling effect  
in the spreading of worms in wireless networks which 
greatly slows down the speed of worm invasion in these networks.

The rest of this paper is organised as follows. In section II we describe
our models of network topology, data communication mechanism, and 
worm spreading in wireless adhoc networks. In section III we present and 
discuss results of our Monte Carlo simulations studies of 
epidemics in these networks for a range of 
device densities and infection rates. 
We close this paper in section IV with conclusions.

\section{Models}
\subsection{Network model}
We consider a collection of nodes distributed in a two
dimensional plane which communicate using short-range 
radio transmissions. The received radio signal 
strength at a device $j$ resulting from  a transmission by a device $i$
decays with the distance between the sender and the receiver due 
to a combination of free-space attenuation and fading effects.
Phenomenologically this effect 
is described using the so-called pathloss model \cite{pathloss}  which states that 
the mean value of the signal power at a receiving device $j$ is related 
to the signal power of the transmitting node $i$  via the following 
equation:
\be
P^{ij}=\frac{P^i}{cr_{ij}^\alpha}.
\ee
In the above equation 
$r_{ij}$ is the Euclidean distance between node $i$ and node $j$, 
$P^{i}$ and $P^{ij}$ are the transmit power and the received power, 
respectively, and $c$ is a constant whose precise value 
depends on a number of factors including the transmission frequency.
For free space propagation $\alpha=2$, but depending on the specific 
indoor/outdoor propagation scenario it is found empirically 
that $\alpha$ can vary
typically between $2$ and $5$. A data transmission by node $i$ is correctly 
received at node $j$, i.e. $i$ can establish a communication link with 
$j$, provided that: 
\be
\frac{P^{ij}}{\nu}= \frac{P^i/cr_{ij}^\alpha}{\nu} \ge \beta_{th}.  
\ee
In the above equation 
$\beta_{th}$ is an attenuation threshold and $\nu$ is the noise level at 
node $j$.

Condition (2) translates 
into a maximum transmission range for node $i$:
\be
r_t^i= \left( \frac{P^i}{c \beta_{th} \nu} \right)^{1/\alpha},
\ee
such that each device can establish wireless  links with only those 
devices within a circle of radius $r_t^i$. 
A communication graph is then constructed 
by creating an edge between node $i$ and all other 
nodes in the plane that are  within the transmission range of $i$, 
and repeating this procedure for all nodes in the network.
In general wireless devices may use different transmit powers 
such that the existence of a wireless link from $i$ to $j$ does not imply 
that a link from $j$ to $i$ also exists. Consequently the resulting 
communication graph is {\it directed}. 
Assuming, however, that all devices use the same transmit power $P$, 
and a corresponding transmission range $r_t$,
the topology of the resulting network can be described as a 
two dimensional random geometric graph (RGG) \cite{rgg1, rgg2}.
Random geometric graphs have been used extensively in the study of 
continuum percolation and more recently 
for modelling wireless adhoc networks \cite{geiner1, geiner2, geiner3, geiner4,
reinel}.
Like  Erd\H os-R\'enyi random graphs (RG) \cite{bollobas}, these graphs 
have a Poisson degree distribution,
$P(k)$, which peaks at an average value $\langle k \rangle$ and 
shows small fluctuations around $\kv$.
However, other 
properties of a RGG are radically different from a
Erd\H os-R\'enyi random  graph. Most notably, these networks are characterised by a 
large cluster coefficient, $C=0.59$,  which is a purely geometric
quantity independent of both node density and $\kv$ \cite{rgg2,geiner1}. 
Furthermore, it has been shown numerically that 
the critical connectivity in these networks is at 
$\kv=4.52$ \cite{rgg2}, which is much higher than the well-known $\kv=1$ value
in RG.

\subsection{Medium access control}
In WiFi networks access to the available frequency channels 
is controlled by a coordination mechanism called 
the Medium Access Control (MAC) \cite{mac}. 
The function of the MAC is to ensure interference-free 
wireless transmissions of data packets in the network. This is achieved 
by scheduling in time the transmissions of nearby devices in such a way that 
devices whose radio transmissions may interfere with each other do not
get access to the wireless channel at the same time. 
The presence of the MAC introduces novel spatio-temporal correlations 
in the dynamics of data communications in these networks which are absent 
in the Internet communications.

The MAC protocol used by  WiFi-based wireless devices follows the 
IEEE 802.11 standard \cite{mac}, which specifies a set of
rules that enable nearby devices coordinate their transmissions 
in a distributed manner. The IEEE 802.11 MAC is a   
highly complex protocol and we do not attempt 
to fully model this protocol. 
Instead we focus on the most relevant aspect of this protocol, the so-called 
listen-before-talk (LBT) rule. This rule  dictates that each device should 
check the occupancy of the wireless medium before starting a data transmission 
and refrain from  transmitting if it senses that the medium is busy. 
The precise implementation of the LBT algorithm will be discussed in
section II.D.

\subsection{Worm propagation model}
Worms are stand-alone computer viruses which use networks for 
their spreading among computing devices. Consequently, 
computer worms can propagate automatically from device to device, 
in contrast to other 
types of virus which require some form of user involvement for their 
spreading. 

Several previous studies have analysed and modelled the propagation of 
computer worms on the Internet.
Most contemporary Internet worms work as follows \cite{book-virus}.
When a computer worm is fired into the Internet, it scans the 
IP (Internet Protocol) addresses and  
sends a probe to infect the corresponding machines.
When a vulnerable machine becomes infected by such a probe, it 
begins running the worm and tries to infect 
other machines. A patch, which repairs the security holes of the
machine, is used to defend against worms. When an infected or
vulnerable machine is patched against a  worm, it becomes immune to  
that worm.  There are  several different scanning mechanisms that
worms deploy. Two main mechanisms are random  scanning and local 
subnet scanning. In random scanning an infected computer scans the 
entire IP address space and selects its targets randomly from this
space. In local scanning the worm scans the nearby targets
(e.g. machines on the same subnet) with a higher probability. Many 
recent worms, such as Code Red v2 have used localised scanning.

The above mechanisms require that
both the infected and the vulnerable nodes are connected to the Internet and 
rely on the IP routing mechanism  for worm delivery.
However, it is well-known that point-to-point routing 
of data packets in wireless adhoc networks could be problematic due 
to the highly dynamic nature of these networks. A much more robust mechanism for 
disseminating packets in such networks is by multihop forwarding in which 
a packet propagates in the network by broadcast radio transmissions from 
device to device, without the need for any routing mechanism or Internet 
connectivity. This  mechanism shows interesting analogies with the 
way airborne diseases spread in  populations and has been exploited in 
a recent worm attack on Bluetooth-enabled  smartphones \cite{sciam}. 
We assume therefore that worms targeting these networks will utilise 
multihop broadcasts as their primary method of propagation.
With respect to an attacking  worm, we assume that nodes in the network to 
be in one of the  following three states: vulnerable,
infected, or  immune. Infected nodes try to transmit the worm to their
neighbours at every possible opportunity.
Vulnerable nodes can become infected at a rate 
$\lambda$ when they receive a transmission containing a
copy  of the worm  from an infected neighbour. Finally, infected 
nodes get patched and become immune to the worm 
at a rate $\delta$. We denote by $S(t)$, $I(t)$ and $R(t)$ the 
population of vulnerable, infected and immune nodes, respectively.

\subsection{Implementation}
In our simulations we  have implemented the above model of worm spreading 
in wireless adhoc networks in the following way. 
At each timestep of simulations we create a randomly ordered list 
of the infected nodes in the network at that timestep.
The first node on the list then gets access 
to the wireless 
channel and is allowed to transmit the worm. All other infected nodes 
that are within the transmission range of this node 
are eliminated from the list as their transmission may cause 
interference to that node, and is therefore blocked by the LBT rule. 
This procedure is repeated for the remaining 
nodes until the list is reduced to a set of 
non-interfering infected nodes which can transmit the worm at that same 
timestep.

Subsequently, all infected nodes which are on the above list go through a broadcast round in which they transmit the worm to their neighbours. Finally, 
all infected nodes (i.e. both those who were able to transmit the worm 
and those whose transmissions were blocked by the MAC protocol) go through a 
patching round in which they may become immune with probability 
$\delta$.

\section{Simulation studies}
We simulated the propagation of worms in wireless adhoc 
networks comprising $N$  devices spread in a $L^2=1000\times 1000$ $m^2$ 
area.  The transmission range of all devices was set at   
$50$ m, which is somewhere between the typical minimum ($30$ m) and  
maximum ($100$ m) range of the WiFi systems.  
In order to investigate the impact of device density 
we performed our simulations for a range of densities, 
corresponding to $N=4000,6000,8000,10000,20000$.

For a given density, nodes were distributed  randomly and uniformly 
in the simulation cell. The resulting RGG networks were constructed 
following the prescription of Sec. II. A, and  periodic 
boundary conditions were used in order to reduce finite-size effects. 
We verified numerically that all the networks considered were connected, and
their  degree distributions were well-described by 
the Poisson distribution:
\be
P(k)=e^{-\langle k \rangle}\frac{\langle k \rangle ^k}{k!},
\ee
with  the average degree,$\kv$, given by: 
\be
\kv=\pi r_t^2\rho,
\ee
where $\rho=N/L^2$ is the device density.

The spreading dynamics was simulated on top of the above networks 
using  Monte Carlo simulations. Each Monte Carlo run starts
by infecting a single randomly chosen node and proceeds 
following  the rules described in sections II.C and II.D until 
the epidemic dies out (i.e. no infected node is left in the network).
We typically average our results over $500$ Monte Carlo runs. 
Furthermore, the results were also averaged over simulations
starting from at least $5$ different initial infected seeds.
Since the time scale of the epidemic spreading depends only on the ratio
$\lambda/\delta$, 
rather than $\lambda$ and $\delta$ separately, without the loss of 
generality we set the patching rate at $\delta=1$ and performed our simulations 
for a range of values of the infection rate, $\lambda$. 

In order to investigate the impact of the MAC on worm epidemics all simulations 
were performed both in the presence and in the absence of this mechanism.
The latter case corresponds to an idealised scenario where nearby devices can communicate with each other without causing harmful interference, for example 
by using non-overlapping frequency channels \cite{foot1}, 
and maps the dynamics 
of worm spreading  onto the standard susceptible-infected-removed 
(SIR) epidemic model. Finally, as a point of reference, 
we also performed simulation 
studies of the SIR model on a set of Erd\H os-R\'enyi 
random graphs
which were constructed such that their degree distribution  
virtually coincided with that of our RGG networks. In the following 
we shall refer to our full simulations as RGG+MAC 
while simulations in the absence of MAC will be labelled as RGG and 
those performed on random graphs as RG. For future reference 
we note that the SIR epidemic on random graphs roughly mimics 
the spread of Internet worms via random scanning, and is  
well-described by the mean-field theory. 

\begin{figure}[t]
 \epsfig{file=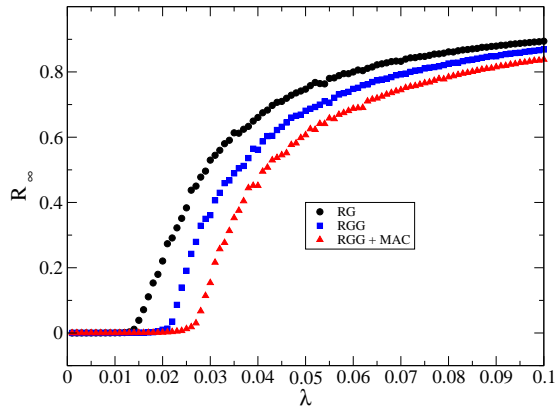, width=2.5in, angle=-90, clip=1}
 \caption{The epidemic prevalence, $R_\infty$, is shown as a function of
 the infection rate $\lambda$ for a wireless adhoc network consisting of 
$N=10000$ nodes, 
both in the absence and presence of the MAC mechanism. Also shown is the result 
for a random graph network with the same number of nodes and the same 
degree distribution as the wireless adhoc network.}
\end{figure}

\begin{figure}[h] 
\epsfig{file=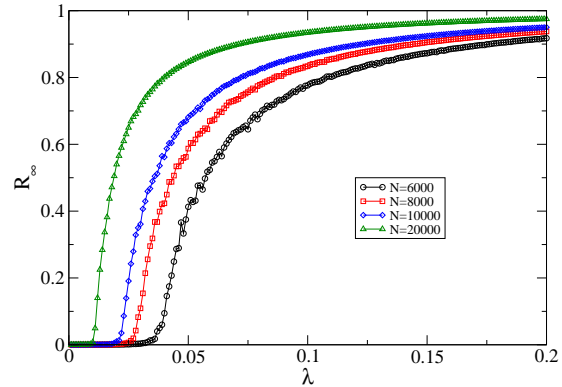,width=2.5in, angle=-90, clip=1}
 \caption{The epidemic prevalence, $R_{\infty}$, as a function of 
the infection rate  $\lambda$ is shown for wireless adhoc networks comprising
    $N=6000,8000,10000,20000 $ nodes, respectively. Results are shown 
for simulations performed in the absence of the MAC mechanism.}
\end{figure}

\begin{figure}[h] 
\epsfig{file=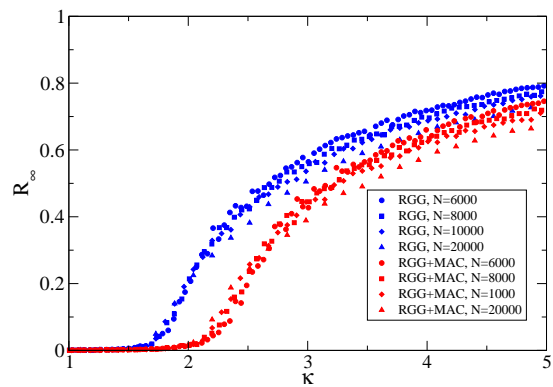,width=2.5in, angle=-90, clip=1}
 \caption{Collapse plots of $R_{\infty}$ vs. 
$\kappa=\lambda \kv$ is shown for wireless adhoc networks comprising 
$N=6000, 8000, 10000, 20000$ nodes, both in the presence and absence 
of the MAC mechanism.}
\end{figure}

\subsection{Prevalence and epidemic threshold}
A key quantity in the study of epidemics in  networks is the epidemic 
prevalence. For the SIR-type epidemics this quantity is 
defined as $R_{\infty}=\lim_{t\rightarrow \infty}R(t)/N$ \cite{sir_yamir}.
In Fig. 1 prevalence as a function of $\lambda$
is shown as obtained from our simulations of, respectively, 
RG, RGG and RGG+MAC networks comprising $N=10000$ nodes. 
It can be seen that in all these networks  $R_{\infty}$ exhibits a critical 
threshold $\lambda_c$ below which a worm cannot spread in the
network and above which it infects a finite fraction of the nodes.
However, the epidemic threshold corresponding to RGG  is at a
considerably higher value than that of RG,
despite the fact that the degree distributions of these two 
networks are identical. Furthermore, it can be seen that the inclusion 
of the MAC mechanism results in an increase in 
the value of the epidemic threshold in RGG, shifting the position of 
$\lambda_c$ even further away from the RG value. 
Our computed epidemic thresholds for the above networks are  
$\lambda=0.0140, 0.0210,0.0265$ for RG, RGG and RGG+MAC, respectively.
The mean-field theory, which in the infinite system size limit 
becomes exact for the RG network, predicts an
epidemic threshold at $\lambda_c=1/\kv$ \cite{sir_yamir,boguna}
, yielding $\lambda_c=0.0127$ for 
the above networks. This is in good agreement with our Monte Carlo 
result for the RG network, indicating that the  
above differences between $\lambda_c$ in RG, RGG and RGG+MAC 
are not due to finite-size effects (or statistical fluctuations) 
but are caused by a combination of topological and dynamic correlations 
in our wireless networks, which are absent in RG. 

Qualitatively, we can understand the above results by noting that due
to spatial ordering in random geometric graphs,  
the epidemic state of a node in a RGG
is  strongly correlated to the state of its neighbours. 
These correlations reduce the so-called reproductive rate of the
epidemic, the average 
number of new infections that can be produced by an infective node, below 
the value predicted by the mean-field approximation, hence increasing $\lambda_c$ above the mean-field value.
The presence of the MAC introduces additional 
temporal correlations between the transmission times of adjacent 
infective nodes which further reduce the reproductive rate, hence 
further increasing $\lambda_c$ above the mean-field value.
It can be seen from Fig.1 that this latter mechanism not only 
affects $\lambda_c$ but also results in a significant reduction 
in the epidemic prevalence.

Next we investigate the impact of node density, $\rho$,
on the behaviour of epidemic prevalence in our networks. In Fig. 2
we plot  $R_{\infty}$ as function of $\lambda$  for different 
device densities. Results are shown only for RGG but they show a 
similar behaviour for RGG+MAC. It can be seen that for all densities 
considered the prevalence shows a critical behaviour. However, the 
position of $\lambda_c$ decreases monotonically with increasing 
density, i.e. the worm epidemic is more  successful in invading the
network when the density of devices
is high and less so when  density is low.
In order to better understand the density dependent 
behaviour of $R_{\infty}$  we plot in Fig. 3 this quantity 
as function of $\kappa=\lambda\kv=\lambda \pi \rho r_t^2$.
It can be seen that for RGG there is a good collapse of curves 
in an extended region around the threshold. This  indicates  
that in this region the prevalence of the SIR model 
on RGG is well-described by the scaling relation 
$R_{\infty}(\lambda,\rho) = f(\kappa)$ (For RGG+MAC this scaling holds 
only approximately).
In particular, we find that the epidemic threshold itself can 
be written as: 
\be
\lambda_c= 
\frac{\kappa_c}{\langle k \rangle}= \frac{\kappa_c}{\pi\rho r_t^2},
\ee
with $\kappa_c=1.50$, a correction to the  mean-field model resulting 
from spatial correlations in RGG. Since the cluster 
coefficient, $C$, is a measure of correlations in a network, we note that 
the value of $\kappa$ is in fact very close to $1/C$ indicating that 
the departure from the mean-field model is possibly 
controlled by this quantity.

Next, we investigate the dependence of the epidemic prevalence on
device density. In Fig. 4 this quantity is plotted as a function of 
$N$ and for $\lambda=0.1$, both in the presence of MAC and when this mechanism is switched off.
It can be seen that in both cases $R_{\infty}$ increases monotonically 
with increasing node density. Furthermore,
for all values of $N$ the curve corresponding to RGG+MAC lies 
below that of RGG. However, the gap between the two curves decreases as 
$N$ is increased, indicating that the 
impact of MAC on $R_{\infty}$ becomes less significant at high densities.

\subsection{Spreading dynamics}
Finally, we discuss the propagation dynamics of worms in our networks. 
Fig. 5 displays, as an example, time evolution of the total 
fraction of infected nodes, $I(t)/N$,
in the $N=10000$ node network and for $\lambda=0.1$. For comparison we 
have also plotted the result obtained for the corresponding random graph 
network. As can be seen from Fig. 5 worm spreading on the RGG network 
takes place at a much slower pace than on the RG network. 
In particular, the initial growth of the epidemic on RGG
is much slower than the exponential growth seen for RG, which is a hallmark of 
mean-field models \cite{bart} and is also observed in epidemics in 
the Internet \cite{worm2}.

The above slow growth  of the epidemic on the RGG is a purely topological 
effect, which we attribute to a combination of spatial correlations and 
high clustering in this network. As can be seen from Fig. 5
switching on the MAC protocol in RGG
slows down the epidemic on this network even further.
This effect, which we call {\it self-throttling},
is caused by 
temporal correlations in the spreading dynamics introduced by the MAC and 
has also been observed in a recent study of IP-based  
worm spreading in mobile adhoc networks \cite{wi-worm}.
It results because adjacent infective devices compete with each other 
in accessing the shared wireless medium, hence effectively blocking each 
other's broadcasts and slowing down the overall progress of 
the epidemic.
In Fig. 6 we display our result for $R(t)/N$ in the $N=10000$ networks, 
which further  demonstrates the great impact of spatial and temporal 
correlations on the spreading process in our wireless networks.

Next we discuss the impact of device density on the speed of worm propagation.
As an indicator of the speed we use 
the position of the peak in $I(t)/N$, which we call $T_{max}$.  
This quantity is plotted in Fig. 7 as a function of $N$, both in the 
absence and presence of MAC. It can be seen that both curves 
show a monotonic decrease with increasing node density.
Furthermore, we see that the self-throttling 
mechanism is most effective in slowing down the epidemic at 
the lowest density, where we observe a $\sim 40\%$ increase in $T_{max}$ when
MAC is switched on. As density increases the gap
between the two curves becomes initially smaller before settling down at 
higher densities.

\begin{figure} 
\epsfig{file=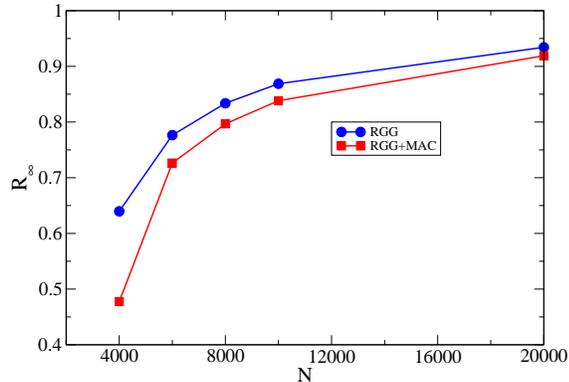, width=2.5in, angle=-90, clip=1}
\caption{The epidemic prevalence in a wireless adhoc network is 
 plotted as a function of the number of devices in the network. 
 Results are shown of simulations 
performed in the absence (circles) and presence of the MAC (squares).}
\end{figure}

\begin{figure}[h]
\epsfig{file=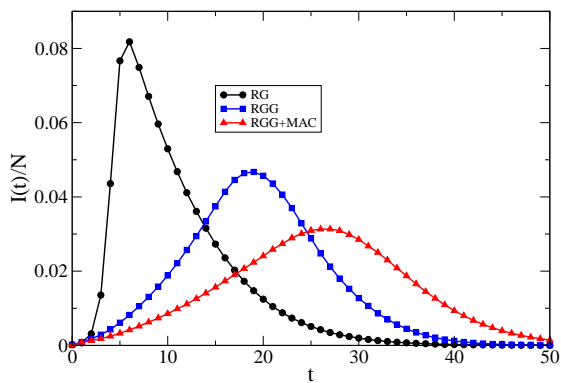,width=2.5in, angle=-90, clip=1} 
\caption{Time evolution of the fraction of infective devices,
$I(t)/N$, is shown for networks consisting of  $N=10000$ nodes.
Results of simulations 
are shown both in the absence (squares) and presence (triangles)
of MAC. Also shown are  the results for the  corresponding 
random graph network (circles).}
\end{figure}

\begin{figure}[h]
\epsfig{file=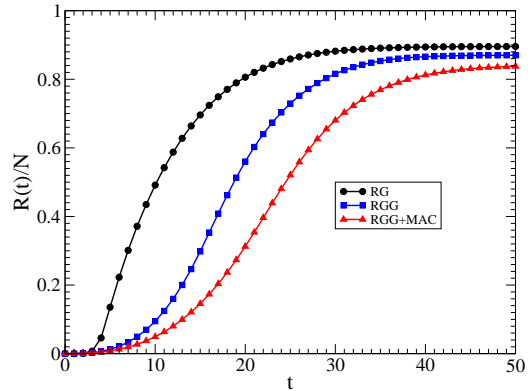,width=2.5in, angle=-90, clip=1} 
\caption{Time evolution of the fraction of immunised devices,
$R(t)/N$, is shown for the same networks as in Fig. 5.}
\end{figure}

\begin{figure}[h] 
\epsfig{file=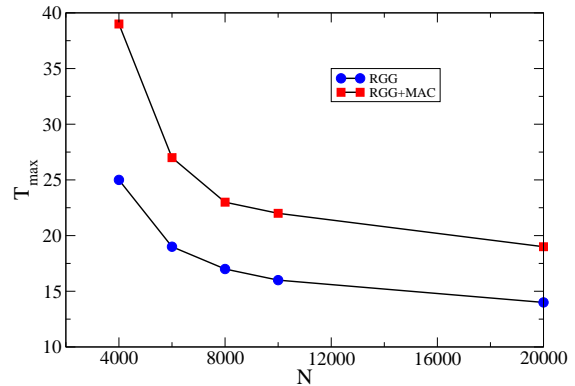,width=2.5in, angle=-90, clip=1}
 \caption{ The position of the epidemic peak, $T_{max}$, in RGG networks
   is plotted as function of $N$. Results are shown when the MAC 
   is switched off (circles) and when it is switched on (squares).}
\end{figure}

\section{CONCLUSIONS}
In this paper we introduced a model for the propagation of a new 
class of  computer worms which specifically target wireless computing devices.
Using extensive Monte Carlo simulations we investigated the 
epidemic spreading of such worms in WiFi-based wireless adhoc networks.
We incorporated the spatial topology 
of these networks via a random geometric graph model, and also 
took into account the impact of the Medium Access Control on 
wireless data communications in these networks.

Our  studies show that worm epidemics 
in wireless adhoc networks are greatly different from the previously 
studied epidemics in the Internet. The epidemic threshold was found to be 
density-dependent and for all densities considered significantly 
higher than the value predicted by the mean-field 
theory. Furthermore, the initial growth of the epidemic
was found to be significantly slower than the exponential growth 
observed in Internet epidemics and predicted by the mean-field theory.
We showed that these differences were due to a combination of spatial 
and temporal correlations which are inherent to wireless data networks.
Our study also revealed the presence of a self-throttling mechanism which 
results from a competition between adjacent infected devices for 
access to the shared wireless medium. This mechanism greatly reduces the 
speed of worm propagation and the risk of large-scale 
worm epidemics in these networks.

An understanding of the propagation characteristics of 
worm attacks on wireless networks is of great importance for
the design of effective detection and prevention strategies for 
these networks. The work presented in this paper is a first 
step in this direction and, we hope, will inspire 
future empirical and theoretical investigations.
From the perspective of complex network theory, our work presents
an extensive study of epidemic spreading in 
random geometric graphs, and highlights the important 
role that spatial correlations play in dynamic processes on these 
and other spatial networks.

\begin{acknowledgments}
M.N. acknowledges support from the Royal Society
through an Industry Fellowship and thanks the Centre for 
Computational Science at UCL for hospitality.
\end{acknowledgments}


\begin{thebibliography}{40}
\bibitem{foot1} 
This might be
the case, for example, in worm attacks on Bluetooth networks.
\end{thebibliography}
\end{document}